\documentclass[12pt]{article}

\usepackage{amssymb}
\usepackage{amsmath}
\usepackage{latexsym}
\usepackage{graphicx}

\catcode`@=11
\renewcommand{\section}{\@startsection{section}{1}{0pt}{\medskipamount}
{\medskipamount}{\large\bf}}
\numberwithin{equation}{section}
\catcode`@=12

\def\beq{\begin{eqnarray}}    
\def\eeq{\end{eqnarray}}      

\def\ln{\,\mbox{ln}\,}                  
\def\det{\,\mbox{det}\,}                

\newcommand{\R}{{\mathbb R}}

\def\im{\textrm{i}}
\def\eu{\textrm{e}}
\def\diff{\textrm{d}}
\def\sfrac#1#2{{\textstyle\frac{#1}{#2}}}
\def\={\ =\ }

\begin{document}

\begin{titlepage}
\setcounter{page}{0}
\begin{flushright}
ITP--UH--10/13
\end{flushright}

\vskip 1.0cm

\begin{center}

{\LARGE\bf Gribov horizon beyond the Landau gauge}

\vspace{18mm}

{\Large Peter M. Lavrov$\,{}^{\dag}$$^{}\footnote{E-mail:
lavrov@tspu.edu.ru}$\quad
and  \quad
Olaf Lechtenfeld$\,{}^{\ddag}$
$^{}\footnote{E-mail: lechtenf@itp.uni-hannover.de}$
}

\vspace{8mm}

\noindent ${}^{\dag}${\em
Tomsk State Pedagogical University,\\
Kievskaya St.\ 60, 634061 Tomsk, Russia}

\vspace{4mm}

\noindent ${}^{\ddag}${\em
Institut f\"ur Theoretische Physik and Riemann Center for Geometry and Physics,\\
Leibniz Universit\"at Hannover,
Appelstrasse 2, 30167 Hannover, Germany}

\vspace{20mm}

\begin{abstract}
\noindent
Gribov and Zwanziger proposed a modification of Yang-Mills theory
in order to cure the Gribov copy problem. We employ field-dependent
BRST transformations to generalize the Gribov--Zwanziger model
from the Landau gauge to general $R_\xi$ gauges.
The Gribov horizon functional is presented in explicit form,
in both the non-local and local variants.
Finally, we show how to reach any given gauge from the Landau one.
\end{abstract}

\end{center}

\vfill


\end{titlepage}


\section{Introduction and summary}

\noindent
It is long known that the covariant quantization of Yang-Mills theory
is beset by the Gribov problem: the existence of infinitely many discrete
gauge copies even after gauge fixing~\cite{Gribov}. A natural remedy,
suppressing the field integration outside the Gribov horizon, is accomplished by
adding to the action a Gribov horizon functional~\cite{Gribov}--\cite{Zwanziger4}.
The latter, however, is not BRST invariant and usually chosen in the Landau gauge.
For a better understanding of its effect on the gauge variance of Greens functions,
a knowledge of the horizon functional in other gauges is desirable~\cite{LLR}.

Recently, we have discovered an explicit way to change the gauge in
Faddeev-Popov quantization by effecting a suitable field-dependent
BRST transformation~\cite{LL}.\footnote{
An analogous formula had been derived differently in~\cite{Mandal}.}
Here, we utilize this strategy to define
horizon functionals for the non-local and local forms of the Gribov--Zwanziger
model in any $R_\xi$ gauge. At the end of the paper, we present the
the horizon functional in an arbitrary gauge.
\\

\section{Yang-Mills theory with Gribov horizon}

\noindent
Yang-Mills theory with gauge group SU$(n)$ in $d$ spacetime dimensions
features gauge potentials $A^a_{\mu}(x)$ with $a=1,\ldots,n^2{-}1$
and $\mu=0,1,\ldots,d{-}1$. The classical action has the standard form
\beq
S_0(A) \= -\sfrac14\int\!\diff^d x\ F_{\mu\nu}^{a}F^{\mu\nu{}a}
\qquad\textrm{with}\qquad
F^a_{\mu\nu}\=\partial_{\mu}A^a_{\nu}-\partial_{\nu}A^a_{\mu}+
f^{abc}A^b_{\mu}A^c_{\nu}\ , \label{clYM}
\eeq
where
$f^{abc}$ denote the (totally antisymmetric)
structure constants of the Lie algebra~$su(n)$. The action
(\ref{clYM}) is invariant under the gauge transformations
\beq
\delta A^a_{\mu}\=D^{ab}_{\mu}\xi^b \qquad\textrm{with}\qquad
D^{ab}_{\mu}\=\delta^{ab}\partial_{\mu}+f^{acb}A^c_{\mu}\ .
\eeq
The BRST formulation of the quantum theory extends the field content to
\beq
\bigl\{\phi^A\bigr\}\=\bigl\{A^a_{\mu}, B^a, C^a, {\bar C}^a\bigr\}
\eeq
by adding the Nakanishi-Lautrup auxiliary fields as well as the
Faddeev-Popov ghost and antighost fields, in the order above.
The Grassmann parities~$\varepsilon$ and ghost numbers~$gh$ are
\beq
\begin{aligned}
&\varepsilon(C^a)=\varepsilon({\bar C}^a)=1\ ,\quad
\varepsilon(A^a_\mu)=\varepsilon(B^a)=0\ ,\\
& gh(A^a_\mu)=gh(B^a)=0, \quad gh(C^a)=-gh({\bar C}^a)=1\ .
\end{aligned}
\eeq

In DeWitt notation~\cite{DeWitt},
the quantum action \`a la Faddeev and Popov~\cite{FP} takes the form
\beq
\label{ExtYMBV}
S(\phi)\= S_0(A)\ +\ {\bar C}^a K^{ab}(A)\,C^b\ +\ \chi^a(A)\,B^a\ ,
\eeq
with the Faddeev-Popov operator
\beq
\label{FPmat}
K^{ab}(A)\=\frac{\delta\chi^a(A)}{\delta A^c_{\mu}}D^{cb}_{\mu}\=
\partial^{\mu}D_{\mu}^{ab}\=
\delta^{ab}\partial^{\mu}\partial_{\mu}+ f^{acb}A^c_{\mu}\partial^{\mu}
\eeq
for the gauge-fixing functions $\chi^a$ of the Landau gauge,
\beq
\chi^a(A) \= \partial^{\mu}\!A^a_{\mu}\ .
\eeq
The action~(\ref{ExtYMBV}) is invariant
under the BRST transformation~\cite{brs,t}
\beq
\label{BRSTtr}
\delta_{\lambda} A_{\mu}^{a} = D^{ab}_{\mu}C^b\lambda\ ,\quad
\delta_{\lambda} \bar{C}{}^a = B^a\lambda\ ,\quad
\delta_{\lambda} B^a = 0\ ,\quad
\delta_{\lambda} C^a = \sfrac12 f^{abc}C^bC^c\lambda
\label{BRSTGZred}
\eeq
where $\lambda$ is an odd constant Grassmann parameter.
Introducing the Slavnov variation $sX$ of any functional $X(\phi)$ via
\beq
\delta_{\lambda}X(\phi)\=\big(sX(\phi)\big)\,\lambda
\qquad\textrm{so that}\qquad
sX(\phi)\=\frac{\delta X(\phi)}{\delta\phi^A}R^A(\phi)
\eeq
with the notation
\beq
\label{RA}
\bigl\{R^A(\phi)\bigr\}\=
\bigl\{D^{ab}_{\mu}C^b,\; 0\;,\; \sfrac12 f^{abc}C^bC^c,\; B^a\bigr\}
\qquad\textrm{and}\quad \varepsilon(R^A(\phi))=\varepsilon_A{+}1\ ,
\eeq
the action (\ref{ExtYMBV}) can be written in the compact form
\beq
\label{Scompact}
S(\phi)\=S_0(A)\ +\ s\psi(\phi)\ ,
\eeq
where $\psi(\phi)$ denotes the the associated fermionic
gauge-fixing functional (in the Landau gauge),
\beq
\label{psi}
\psi(\phi)\={\bar C}^a\chi^a(A)\={\bar C}^a\,\partial^{\mu}\!A^a_{\mu}\ .
\eeq

The Gribov horizon \cite{Gribov} in the Landau gauge can be taken
into account by adding to the action~(\ref{Scompact}) the non-local
horizon functional
\beq \label{FuncM}
M(A)\=\gamma^2\,f^{abc}A^b_{\mu}(K^{-1})^{ad}f^{dec}A^{e\mu}
\ +\ \gamma^2\,d(n^2{-}1)\ ,
\eeq
where $K^{-1}$ inverts the (matrix-valued) Faddeev-Popov operator $K^{ab}(A)$
of~(\ref{FPmat}) and $\gamma\in\R$ is the so-called thermodynamic or Gribov
parameter~\cite{Zwanziger1,Zwanziger2}.
The effective action of the Gribov--Zwanziger model,
\beq
\label{GZact} S_M(\phi)\=S(\phi)\ +\ M(A)\=S_0(A)\ +\ s\psi(\phi)\ +\ M(A)\ ,
\eeq
is not BRST invariant because
\beq
\label{sM}
sM(A,C)\=\gamma^2f^{abc}f^{cde}\bigl[2D^{bq}_{\mu}C^q(K^{-1})^{ad}-
f^{mpn}A^b_{\mu}(K^{-1})^{am}K^{pq}C^q(K^{-1})^{nd}\bigr]A^{e\mu}\ \neq\ 0\ .
\eeq
In \cite{LLR}, we have investigated the resulting gauge dependence of the vacuum
functional, assuming the existence of a horizon functional beyond the
Landau gauge. With the help of recent results~\cite{LL} (see also~\cite{Mandal}),
we now verify this assumption and propose an explicit form for such a functional
in general $R_\xi$ gauges.
\\

\section{Gribov horizon in $R_{\xi}$ gauges}

\noindent The vacuum functional for the Gribov--Zwanziger model is
given by a functional integral,
\beq \label{vint} Z\=\int\!{\cal
D}\phi\ \exp\Big\{\frac{\im}{\hbar}\big( S_0(A)+
s\psi(\phi)+M(A)\big)\Big\} \ .
\eeq
Let us perform a change of
variables which amounts to a particular field-dependent BRST
transformation,
\beq \label{chv} \phi^A\quad \longmapsto \quad
\phi^A+(s\phi^A)\Lambda_\xi(\phi) \qquad\textrm{with}\qquad
\Lambda_\xi(\phi)\={\bar C}^a B^a\,(B^2)^{-1}
\big(\exp\big\{\sfrac{\xi}{2\im\hbar}B^2\big\}-1\big)\ ,
\eeq
where $B^2=B^aB^a$. Taking
into account the Jacobian and using\
$\ln(1+s\Lambda_\xi)=\sfrac{\xi}{2\im\hbar}B^2$, the vacuum
functional then reads~\cite{LL}
\beq \label{vintt} Z\=\int\!{\cal D}
\phi\ \exp\Big\{\frac{\im}{\hbar}\big(
S_0(A)+s\psi_\xi(\phi)+M_\xi(\phi)\big)\Big\}\ ,
\eeq
with a shifted
fermionic gauge-fixing functional and a modified horizon functional,
\beq \label{GZhR} \psi_\xi(\phi)\={\bar
C}^a\big(\partial^{\mu}\!A^a_{\mu}+\sfrac{\xi}{2}B^a\big)
\qquad\textrm{and}\qquad
M_\xi(\phi)\=M(A)+\big(sM(A,C)\big)\Lambda_\xi(\phi)\ ,
\eeq
respectively. The explicit expression for $sM(A,C)$ is given
in~(\ref{sM}).

We have moved away from the Landau gauge and reached a general $R_\xi$ gauge.
Therefore, we propose
\beq
\label{ExGh}
M_\xi(\phi)&=&\gamma^2\,f^{abc}A^b_{\mu}(K^{-1})^{ad}f^{dec}A^{e\mu}
\ +\ \gamma^2\,d(n^2{-}1) \\[6pt] \nonumber
+\ \gamma^2\,f^{abc}\!\!\!\!\!\!\!&&\!\!\!\!\!\!\!\!f^{cde}
\big[2D^{bq}_{\mu}C^q(K^{-1})^{ad}
-f^{mpn}A^b_{\mu}(K^{-1})^{am}K^{pq}C^q(K^{-1})^{nd}\big] A^{e\mu}
{\bar C}^\ell B^\ell(B^2)^{-1}\big(\eu^{\frac{\xi}{2\im\hbar}B^2}{-}1\big)
\eeq
as the explicit form for the horizon functional in a general $R_\xi$ gauge.
Under further BRST transformations, its Slavnov variation is
\beq
sM_\xi \= sM(A,C)\bigl[ 1-s\Lambda_\xi(\phi)\bigr]\ .
\eeq
In linear approximation in~$\xi$ we have \
$\Lambda_\xi(\phi)=\sfrac{\xi}{2i\hbar}{\bar C}^a\!B^a$ \ and get
\beq
M_\xi \= M(A) + \sfrac{\xi\gamma^2}{2\im\hbar} f^{abc}f^{cde}
\big[2D^{bq}_{\mu}C^q(K^{-1})^{ad} -f^{mpn}A^b_{\mu}
(K^{-1})^{am}K^{pq}C^q(K^{-1})^{nd}\big] A^{e\mu} {\bar C}^\ell B^\ell
\eeq
still depending on all field variables.
For $\xi{=}0$, it smoothly reduces to the Landau-gauge functional, $M_0=M(A)$.
It is important to note that our extension~(\ref{ExGh}) of the Gribov--Zwanziger
model is done in such a way as to render its vacuum functional gauge invariant.
Indeed, since an infinitesimal change of gauge $\delta\psi=\im\hbar\Lambda_\xi$
(in linear approximation) is merely a field redefinition in the path integral,
we have that
\beq
\bigl\langle \delta M(\phi) \bigr\rangle\ +\
\bigl\langle s\,\delta\psi(\phi) \bigr\rangle \=0
\eeq
for expectations values $\langle\cdots\rangle$ in the Gribov--Zwanziger model
defined by~(\ref{vint}).
\\

\section{Gribov--Zwanziger action}

\noindent
Originally, the Gribov--Zwanziger model was presented in the non-local form
(\ref{FuncM}) and (\ref{GZact})~\cite{Gribov,Zwanziger1}. Later, the non-locality was
`resolved' by adding auxiliary field variables~\cite{Zwanziger2,Zwanziger3,Zwanziger4}.
The resulting local action is referred to as the Gribov--Zwanziger action
and takes the form (for details, see~\cite{Kondo})
\beq
\label{SGZ}
S_{GZ}(\Phi)\=S_0(A)\ +\ s\psi(\phi)\ +\
S_{\gamma}(A,\varphi,{\bar\varphi},\omega,{\bar\omega})
\eeq
where
\beq
\label{Sgamma}
S_{\gamma}\={\bar\varphi}^{ac}_\mu K^{ab}\varphi^{\mu bc}\ -\
{\bar\omega}^{ac}_\mu K^{ab}\omega^{\mu bc}\ +\ 2\im\gamma f^{abc}A^{b}_{\mu}
\big(\varphi^{\mu ac}+{\bar\varphi}^{\mu ac}\big)\ +\ \gamma^2\,d(n^2{-}1)
\eeq
represents the horizon functional written in local form for the Landau gauge.
The set of fields has been further enlarged to
\beq
\bigl\{\Phi^{\cal A}\bigr\}\=
\bigl\{\phi^A\,,\,\varphi^{ac}_\mu\,,\,{\bar\varphi}^{ac}_\mu\,,\,
\omega^{ac}_\mu\,,\,{\bar\omega}^{ac}_\mu\bigr\}\ .
\eeq
The fields $\varphi^{ac}_\mu$ and ${\bar\varphi}^{ac}_\mu$ are commuting while
$\omega^{ac}_\mu$ and ${\bar\omega}^{ac}_\mu$ are anticommuting.
The additional fields form BRST doublets~\cite{DSV},
\beq
\begin{aligned}
\delta_{\lambda} \varphi^{ac}_\mu &=\omega^{ac}_\mu\lambda\ ,\qquad\ \,
\delta_{\lambda} {\bar\varphi}^{ac}_\mu=0\ ,\\
\delta_{\lambda} \omega^{ac}_\mu &=0\ ,\qquad\qquad
\delta_{\lambda} {\bar\omega}^{ac}_\mu =-{\bar\varphi}^{ac}_\mu\lambda\ .
\end{aligned}
\eeq The local horizon functional $S_{\gamma}$ is not BRST
invariant, \beq \label{varSg}
sS_{\gamma}\=f^{adb}\big[{\bar\varphi}^{ac}_\mu
K^{de}C^e\varphi^{\mu bc}+ {\bar\omega}^{ac}_\mu
K^{de}C^e\omega^{\mu bc}+ 2\im\gamma\big(
D^{de}_{\mu}C^e(\varphi^{\mu ab}{+}{\bar\varphi}^{\mu ab})+
A^d_{\mu}\omega^{\mu ab}\big)\big]\ \neq\ 0\ .
\eeq
Note that in case $\gamma=0$ the action (\ref{Sgamma}) is reduced to
\beq
S_{\gamma=0}\={\bar\varphi}^{ac}_\mu K^{ab}\varphi^{\mu bc}\ -\
{\bar\omega}^{ac}_\mu K^{ab}\omega^{\mu bc}\ .
\eeq
Then, in the vacuum functional, integration over ${\bar\varphi}$ and $\varphi$ yields
$(\det K)^{-1}$, while integration over ${\bar\omega}$ and $\omega$ reproduces
$\det K$, so that in the configuration space $\{\phi\}$ we recover the Yang-Mills 
vacuum functional.

Like in the previous section, we may move to a general $R_\xi$ gauge by
performing the specific field-dependent BRST transformation~(\ref{chv})
in the vacuum functional integral of the Gribov--Zwanziger model
based on the local action~(\ref{SGZ}). As a result, the action gets modified,
\beq
\label{SGZtr}
S_{GZ}(\Phi)\ \longmapsto\
S_0(A)\ +\ s\psi_\xi(\phi)\ +\ S_{\gamma\xi}(\Phi)
\eeq
where
\beq
\label{exSgR}
\psi_\xi(\phi)={\bar C}^a\big(\partial^{\mu}\!A^a_{\mu}+\sfrac{\xi}{2}B^a\big)
\quad\textrm{and}\quad
S_{\gamma\xi}(\Phi)=S_{\gamma}(A,\varphi,{\bar\varphi},\omega,{\bar\omega})
+\big(sS_{\gamma}(A,C,\varphi,{\bar\varphi},\omega,{\bar\omega})\big)\Lambda_\xi(\phi)\,.
\eeq
We propose this $S_{\gamma\xi}$ together with (\ref{chv}), (\ref{Sgamma})
and~(\ref{varSg}) as the proper extension of the local horizon functional
to a general $R_{\xi}$ gauge.
Its Slavnov variation reads
\beq
sS_{\gamma\xi} \= sS_{\gamma}(A,C,\varphi,{\bar\varphi},\omega,{\bar\omega})
\bigl[1-s\Lambda_\xi(\phi)\bigr]\ .
\eeq
In the limit $\gamma\rightarrow0$ we expect the action~(\ref{SGZtr}) again to
produce the standard Yang-Mills theory. Putting $\gamma=0$ in~(\ref{exSgR}) we get
\beq
S_{\gamma=0,\xi}\={\bar\varphi}^{ac}_\mu D^{ab}_{\xi}\varphi^{\mu bc}\ -\
{\bar\omega}^{ac}_\mu D^{ab}_{\xi}\omega^{\mu bc}
\qquad\textrm{where}\qquad
D^{ab}_{\xi}=K^{ab}+f^{adb}K^{de}C^e\Lambda_{\xi}(\phi)\ .
\eeq
Like at $\xi{=}0$ before, integrating out all auxiliary fields indeed leads back to
the Yang-Mills vacuum functional.

With this information, we may revisit the gauge dependence of Greens functions
proposed in~\cite{LLR}, by taking into account the gauge variation of source
terms to be added in the path integral.
\\

\section{Horizon functional in an arbitrary gauge}

\noindent
Although the $R_\xi$ gauges were easy to reach, they are not the only ones
accessible by our method. In fact, \cite{LL} provides a general formula for
connecting any two gauges in terms of their fermionic gauge-fixing functionals~$\psi$:
To get from a reference gauge~$\psi_0$ to a desired gauge~$\psi$, change the variables
inside the generating functional~$Z(J)$ by a BRST transformation with a field-dependent
parameter
\beq
\begin{aligned}
\Lambda_\psi(\phi) &\= (\psi{-}\psi_0)\bigl(s(\psi{-}\psi_0)\bigr)^{-1}
\bigl( \exp\bigl\{\sfrac{1}{\im\hbar}s(\psi{-}\psi_0)\bigr\}-1 \bigr)\\
&\= \sfrac{1}{\im\hbar}(\psi{-}\psi_0)
\sum_{n=0}^\infty\sfrac{1}{(n+1)!}\bigl(\sfrac{1}{\im\hbar}s(\psi{-}\psi_0)\bigr)^n\ .
\end{aligned}
\eeq
The corresponding change of the horizon functional reads
\beq
M_\psi(\phi) - M_0(\phi) \= \big(sM_0(\phi)\big)\Lambda_\psi(\phi)\ .
\eeq
The Gribov--Zwanziger model can now be studied explicitly in an arbitrary gauge.
\\

\section*{Acknowledgments}
\noindent The authors thank I.L.~Buchbinder, I.V.~Tyutin and D.
Zwanziger for useful discussions. This work was supported by the DFG
grant LE 838/12-1. The work of PML is also supported by the LRSS
grant 224.2012.2,  by the Ministry of Education and Science of
Russian Federation, project 14.B37.21.0774, by the RFBR grant
12-02-00121 and the RFBR-Ukraine grant 13-02-90430. He is grateful
to the Institute of Theoretical Physics at Leibniz University for
hospitality.

\end{document}